\documentclass[fleqn,twoside]{article}
\usepackage{amsmath,enumerate,capt-of,ifthen,calc}

\newcommand{\tmem}[1]{{\em #1\/}}
\newcommand{\tmop}[1]{\ensuremath{\operatorname{#1}}}
\newcommand{\mathd}{\mathrm{d}}
\newcommand{\mathi}{\mathrm{i}}
\newcommand{\mathe}{\mathrm{e}}
\newcommand{\bignone}{}
\newcommand{\tmfloatcontents}{}
\newlength{\tmfloatwidth}
\newcommand{\tmfloat}[5]{
  \renewcommand{\tmfloatcontents}{#4}
  \setlength{\tmfloatwidth}{\widthof{\tmfloatcontents}+1in}
  \ifthenelse{\equal{#2}{small}}
    {\ifthenelse{\lengthtest{\tmfloatwidth > \linewidth}}
      {\setlength{\tmfloatwidth}{\linewidth}}{}}
    {\setlength{\tmfloatwidth}{\linewidth}}  \begin{minipage}[#1]{\tmfloatwidth}
    \begin{center}
      \tmfloatcontents
      \captionof{#3}{#5}
    \end{center}
  \end{minipage}}
\newcommand{\mathpi}{\pi}
\newenvironment{enumeratenumeric}{\begin{enumerate}[1.]}{\end{enumerate}}

\usepackage{espcrc2}

\readRCS
$Id: espcrc2.tex,v 1.2 2004/02/24 11:22:11 spepping Exp $
\title{The $A_5$ and the pion field\thanks{Talk given at QCD05,
    Montpellier, France, July 2005}}

\author{Johannes Hirn\address{IFIC, Departament de F\'\i sica
    Te\`orica, CSIC - Universitat de Val\`encia\\ Edifici d'Instituts de
    Paterna, Apt. Correus 22085, 46071 Val\`encia, Spain\\
{\tt    johannes.hirn@ific.uv.es}}\thanks{Speaker.}%
 and Ver\'onica Sanz\address{CAFPE, Departamento de F\'\i sica
    Te\'orica y del Cosmos, Universidad de Granada\\ Campus de
    Fuentenueva, 18071 Granada, Spain\\ {\tt vsanz@ugr.es}}}
       
\runtitle{Proceedings of QCD05, Montpellier, France, 4-8 July 05}

\begin{document}

\begin{abstract}
In this talk, an $\tmop{SU}\left( N_f \right) \times \tmop{SU}\left( N_f \right)$ Yang-Mills model
with a compact  extra-dimension is
used to describe the spin-1 mesons and pions of massless QCD in the
large-$N_c$. The right 4D symmetry and symmetry-breaking pattern
is produced by imposing appropriate boundary conditions. The
Goldstone boson (GB) fields are constructed using
a Wilson line. We derive the low-energy limit (chiral lagrangian),
discuss $\rho$-meson dominance, sum rules between resonance couplings
and the relation with the QCD high-energy behavior. Finally, we
provide an analytic expression for the two-point function of vector
and axial currents.
\vspace{1pc}
\end{abstract}

\maketitle

\section{Introduction}

Based on \cite{Hirn:2005nr}, we present a model for a sector of large-$N_c$ QCD in the chiral limit: that
of pions and the {\tmem{infinite tower}} of spin-1 mesons. In view of a
leading $1 / N_c$ approximation, we limit ourselves to tree-level. The model
never refers to quarks, but rather to mesons. These
will be described by Kaluza-Klein (KK) excitations of five-dimensional (5D)
fields.

Consider $\tmop{SU} \left( N_f \right) \times \tmop{SU} \left( N_f \right)$
Yang-Mills fields in a compact extra dimension. From the four-dimensional (4D)
point of view, these are seen as two infinite towers of resonances. Imposing appropriate boundary conditions
(BCs) on one boundary will induce spontaneous breaking of the 4D $\tmop{SU} \left(
N_f \right) \times \tmop{SU} \left( N_f \right)$ symmetry acting on the other
boundary (where the standard 4D electroweak interactions are located).

We define the model in Section \ref{model}. We concentrate in Section
\ref{GBs} on the predictions for low-energy physics, i.e. pion interactions.
At higher energies, resonances play the leading role: we consider the
constraints on their interactions in Section \ref{res}. Sum rules for
their couplings are obtained, relying on 5D
gauge structure: they imply a soft high-energy behavior. This allows a partial
matching to QCD. Up to this point, the geometry (5D metric) influences none of
the qualitative features of the model. We discuss $\rho$-meson
dominance in Section \ref{LMD}. In Section \ref{2-point}, we focus on the AdS$_5$
metric and give analytic results for the vector and axial two-point
functions.

\section{The model} \label{model}

Using {\tmem{conformally flat}} coordinates $\mathd s^2 = w \left( z \right)^2
\left( \eta_{\mu \nu} \mathd x^{\mu} \mathd x^{\nu} - \mathd z^2 \right)$, we
consider a compact extra dimension extending from $z = l_0$ ({\tmem{UV
brane}}) to $z = l_1$ ({\tmem{IR brane}}). The standard $\tmop{SU} \left( N_f
\right) \times \tmop{SU} \left( N_f \right)$ Yang-Mills action in 5D reads $M
= ( \mu, 5 )$
\begin{eqnarray}
  \mathcal{S}_{\tmop{YM}} & = & - \frac{1}{4 g_5^2}  \int \mathd^4 x
  \int_0^{l_1} \mathd zw \left( z \right) \eta^{M N} \eta^{R S} \nonumber\\
  & \times & \left\langle L_{M R} L_{N S} + R_{M R} R_{N S} \right\rangle . 
  \label{action}
\end{eqnarray}
Having a larger (i.e. 5D instead of 4D) $\tmop{SU} \left( N_f \right) \times
\tmop{SU} \left( N_f \right)$ symmetry will have important consequences on the
high-energy behavior of the model, see Section \ref{res}. As for now, we
discuss the fate of the elements of the gauge group acting on the two
boundaries, i.e. the BCs.

$\bullet$ {\tmem{UV brane}}: The fields at $z = l_0$ are taken to be classical
sources: this defines a generating functional. The 4D $\tmop{SU} \left( N_f
\right) \times \tmop{SU} \left( N_f \right)$ local symmetry at $z = l_0$ (a
subgroup of the 5D gauge invariance) then guarantees the Ward identities
of a 4D $\tmop{SU} \left( N_f \right) \times \tmop{SU} \left( N_f \right)$
global symmetry.

$\bullet$ {\tmem{IR brane}}: Of the $\tmop{SU} \left( N_f \right) \times
\tmop{SU} \left( N_f \right)$ transformations acting at $z = l_1$, we allow
only the vector subgroup as a symmetry of the whole theory. This is done by
imposing the BCs
\begin{eqnarray}
  R_{\mu} \left( x, z = l_1 \right) - L_{\mu} \left( x, z = l_1 \right) & = &
  0,  \label{BC1}
\end{eqnarray}
\begin{eqnarray}
  R_{5 \mu} \left( x, z = l_1 \right) + L_{5 \mu} \left( x, z = l_1 \right) &
  = & 0 .  \label{BC2}
\end{eqnarray}
\section{Goldstone bosons} \label{GBs}

With the above BCs, the spectrum contains a massless 4D spin-0 mode, related
to the (axial combination of) the fifth component of the gauge fields
{\footnote{This is different from the 5D model of
{\cite{Erlich:2005qh,daRold:2005zs}}, where new bulk scalars mix with the
$A_5$ to yield a zero mode.}}. The corresponding covariant object is obtained
by constructing the following Wilson line
\begin{eqnarray}
  U \left( x \right) & \equiv & \text{P} \left\{ \mathe^{\mathi
  \int_{l_1}^{l_0} \mathd zR_5 \left( x, z \right) \bignone} \right\}
  \nonumber\\
  & \times & \text{P} \left\{ \mathe^{\mathi \int_{l_0}^{l_1} \mathd zL_5
  \left( x, z \right) \bignone} \right\} . 
\end{eqnarray}
$U$ depends on the first four coordinates only. Moreover, of the whole
(5D) group, $U$ is only affected by those transformations acting on the UV
brane, in a manner that is precisely appropriate for GBs of the
4D chiral symmetry defined in $z = l_0$: the breaking is spontaneous.

The interactions of the massless modes (the pions) at low energy can then be
obtained. Paying attention to preserve the 4D chiral symmetries, one performs
the identification with the chiral lagrangian operator by operator. The pion
decay constant can be read off as $1 / f_{\pi}^2 = g_5^2  \int_{l_0}^{l_1}
\frac{\mathd z}{w \left( z \right)}$. The $\mathcal{O}( p^4 )$ low-energy
constants of the $\chi$PT lagrangian can be computed. In addition, the
following relations hold {\tmem{independently of the metric}} (the
first two are familiar from {\cite{Ecker:1989yg}})
\begin{eqnarray}
  L_2  &=&  2 L_1 =- 1/3  L_3, \label{L2}\\
  L_{10} & = & - L_9  .  \label{L9+L10}
\end{eqnarray}

\section{Resonances} \label{res}

Having extracted the massless scalar, we perform redefinitions to eliminate
the fifth components. Also, due to the BCs (\ref{BC1}-\ref{BC2}), the action
will be diagonal in terms of vector and axial combinations of the original
gauge fields. One then follows the standard KK procedure: a 5D field $V_{\mu}
\left( x, z \right)$ is decomposed as a sum of 4D modes $V_{\mu}^{\left( n
\right)} \left( x \right)$, each possessing a profile $\varphi_n^V \left( z
\right)$ in the fifth dimension, i.e. $V_{\mu} \left( x, z \right) = \sum_{n =
1}^{\infty} V^{\left( n \right)}_{\mu} \left( x \right) \varphi^V_n \left( z
\right)$. The wave-functions $\varphi \left( z \right)$ are entirely
determined by the metric, as solutions of $- \partial_z \left( w \left( z
\right) \partial_z \right) \varphi^X_n = M_{X_n}^2 w \left( z \right)
\varphi_n^X$, with BCs deduced from those of Section \ref{model}. The BCs are found to be, for the vector case
$\varphi^V_n \left( z \right) \left|_{z = l_0} = \partial_z \varphi_n^V \left(
z \right) \right|_{z = l_1} = 0$, and for the axial one $\varphi_n^A \left(
z \right) \left|_{z = l_0} = \varphi_n^A \left( z \right) \right|_{z = l_1}
= 0$.

This yields an alternating tower of vector and axial resonances. The masses of
the heavy resonances behave as $M_{V_n, A_n} \sim n$ as a consequence of the
5D Lorentz invariance broken by the finite size of the extra-dimension. This
conflicts with the expectation $M_{V_n, A_n} \sim \sqrt{n}$ from the
quasi-classical hadronic string {\cite{Shifman:2005zn}}.

Couplings of resonances are expressed as bulk integrals over the fifth
coordinate. One derives sum rules
 using
the completeness relation for the KK wave-functions
{\footnote{For definitions and derivations see {\cite{Hirn:2005nr}}, and also {\cite{Sakai:2005yt}}.}}
\begin{eqnarray}
  \sum_{n = 1}^{\infty} f_{V_n} g_{V_n} & = & 2 L_9,  \label{fVgV}\\
  \sum_{n = 1}^{\infty} f_{V_n} g_{V_n} M_{V_n}^2 & = & f_\pi^2, 
  \label{fVgVMV2}\\
  \sum_{n = 1}^{\infty} g_{V_n}^2 & = & 8 L_1,  \label{L1SR}\\
  \sum_n g_{V_n}^2 M_{V_n}^2 & = & \frac{f_\pi^2}{3},  \label{KSFRsum}\\
   \sum_{n = 1}^{\infty} \left( f_{V_n}^2 - f_{A_n}^2 \right) & = & - 4
  L_{10} .  \label{L10}
\end{eqnarray}

The first three relations are reminiscent of those considered in
{\cite{Ecker:1989yg}}, but generalized to infinite sums. They ensure that the
vector form factor (VFF) and GB forward elastic scattering amplitude
respectively satisfy an unsubtracted and once-subtracted dispersion relation,
as expected in QCD. The sum rule (\ref{KSFRsum}) shows that the natural value
for the KSFR II ratio is $3$ here, rather than $2$ {\footnote{A similar result was
noted in {\cite{Son:2003et,daRold:2005zs}}.}}.

\section{$\rho$-meson dominance} \label{LMD}

Before extracting numerical values, we consider the question of $\rho$-meson
dominance. As can be expected in any 5D model, only the light resonances will
have non-negligible overlap integrals with the pion field, since the GBs are
non-local objects in the fifth dimension {\cite{Son:2003et}}. As a consequence,
the sums (\ref{fVgV}-\ref{KSFRsum}) are dominated by the first
resonance, the $\rho$. In Table \ref{TLMD}, we focus on the VFF, which can be
put in the form $F \left( q^2 \right) = \sum_{n = 1}^{\infty} \frac{f_{V_n}
g_{V_n} M_{V_n}^2}{f_\pi^2}  \frac{M_{V_n}^2}{M_{V_n}^2 - q^2} \bignone$: two
metrics are considered, flat space and AdS.

\tmfloat{h}{big}{table}{
\[ \begin{array}{|c|c|c|}
     \hline
     & w \left( z \right) = 1 & w \left( z \right) = \frac{l_0}{z}\\
     \hline
     f_{V_1} g_{V_1}  \frac{M^2_{V_1}}{f_\pi^2} & 0.81 & 1.11\\
     \hline
     f_{V_n} g_{V_n}  \frac{M^2_{V_n}}{f_\pi^2} & \underset{n \gg 1}{\sim}
     \frac{1}{n^2} & \underset{n \gg 1}{\sim} \frac{( - 1 )^n}{\sqrt{n}}\\
     \hline
   \end{array} \]}{Behavior of the quantity $f_{V_n} g_{V_n} M_{V_n}^2 / f_\pi^2$:
compare with the results quoted in {\cite{Shifman:2005zn}}.\label{TLMD}}

\vspace{1pc}

The situation is then the following: independently of the metric, the 5D gauge
structure ensures two properties: soft HE behavior through the sum rules
(\ref{fVgV}-\ref{L1SR}) and approximate saturation of these sums by the first
term ($\rho$-meson dominance). Approximating the sum rules by keeping only the
first term, we would recover exactly the same relations as in
{\cite{Ecker:1989yg}}. As in this reference, we are therefore assured that,
once the two input parameters are matched to $M_{\rho} \simeq 776 \tmop{MeV}$
and $f_{\pi} \simeq 87 \tmop{MeV}$, the predictions for the low-energy
constants and decays of the $\rho$ will  provide good estimates, see Table
\ref{Tnum}. This result {\tmem{does not depend strongly on the chosen
metric}}. On the other hand, the masses of the resonances above the $a_1$ grow
too fast with $n$, as mentioned before.

\tmfloat{ht}{big}{table}{
\[ \begin{array}{|c|c|c|c|c|}
     \hline
     &  w \left( z \right) = 1 & w \left( z \right) = \frac{l_0}{z} &
     \text{Experiment}\\
     \hline
     10^3 L_1 & 0.5 & 0.5 & 0.4 \pm 0.3\\
     \hline
     10^3 L_2 & 1.0 & 1.1 & 1.35 \pm 0.3\\
     \hline
     10^3 L_3 & - 3.1 & - 3.1 & - 3.5 \pm 1.1\\
     \hline
     10^3 L_9 & 5.2 & 6.8 & 6.9 \pm 0.7\\
     \hline
     10^3 L_{10} & - 5.2 & - 6.8 & - 5.5 \pm 0.7\\
     \hline
     \Gamma_{\rho \to e e} & 4.4 & 8.1 & 7.02 \pm
     0.11\\
     \hline
     \Gamma_{\rho \to \pi \pi} & 135 & 135 & 150.3 \pm
     1.6\\
     \hline
     M_{A_1} & 1.6 \times 10^3 & 1.2 \times 10^3 & 1230 \pm 40\\
     \hline
     M_{V_2} & 2.3 \times 10^3 & 1.8 \times 10^3 & 1465 \pm 25\\
     \hline
     M_{V_3} & 3.9 \times 10^3 & 2.8 \times 10^3 & 1688.1 \pm 2.1\\
     \hline
   \end{array} \]
}{Numerical outputs. Experimental values for the $L_i$'s correspond to a
renormalization scale $\mu = M_{\rho}$ {\cite{Bijnens:1994qh}}. Dimensionful quantities are
in MeV, except for $\Gamma_{\rho \rightarrow e e}$ in
KeV.\label{Tnum}}

\section{Two-point functions} \label{2-point}

We particularize to an exactly conformal metric $w \left( z \right) = l_0 /
z$: conformal invariance is then broken only by the presence of the IR brane. The vector and axial two-point
functions contain an infinite number of poles, and can be expressed for
time-like momenta as {\footnote{The limit $l_0 \rightarrow 0$ with
    $l_0/g_5^2$ fixed is implied
here. The divergence in $\lambda$ is absorbed by a source contact term, see
{\cite{Hirn:2005nr}}.}}
\begin{eqnarray}
  \Pi_{V, A} ( q^2 ) & = & - \frac{l_0}{g_5^2}  \left( \log \frac{q^2}{\mu^2}
  + \lambda \mp 2 \mathpi \frac{Y_{0, 1} \left( ql_1 \right)}{J_{0, 1} \left(
  ql_1 \right)} \right),  \label{HEPVA}
\end{eqnarray}
where $\lambda = \log ( \mu^2 l_0^2 / 4 ) + 2 \gamma_E $.

The first term in this expression reproduces the partonic logarithm, as
already emphasized in {\cite{Son:2003et,Erlich:2005qh,daRold:2005zs}}. The
last piece in the correlator (\ref{HEPVA}) comes from the breaking of
conformality produced by (different) BCs on the IR brane. Correspondingly, in
the euclidean $Q^2 = - q^2 > 0$, it is suppressed by an exponential factor
$\mathe^{- 2 Q l_1}$ at high energies. Therefore,  (\ref{HEPVA}) does not reproduce any of
the condensates that appear in the QCD operator product expansion (OPE) {\cite{Shifman:1979bx}}.

The absence of these power corrections (in the difference of vector and axial
two-point functions) translates as an infinite number of generalized Weinberg
sum rules {\cite{Barbieri:2003pr}}: this result will hold as long as the
symmetry-breaking effects are localized at a finite distance from the UV
brane.

Expanding now at low energies $ql_1 \ll 1$, the logarithmic contribution is
 canceled by part of the last term in (\ref{HEPVA}) to end up with an
analytic function in powers of $q l_1$
\begin{eqnarray}
  \frac{g_5^2}{l_0} \Pi_V ( q^2 ) & \simeq & c + \frac{1}{2} l_1^2 q^2 +
  \frac{5}{64} l_1^4 q^4 +\mathcal{O} \left( q^6 \right) . 
\end{eqnarray}
The constant term is the expected one from the chiral expansion: in RS1, we
can verify that $c = 2 \log \left( l_1 / l_0 \right) = - 4 L_{10} - 8 H_1$.

We have seen that the result (\ref{HEPVA}) lacked power corrections to
reproduce QCD results. Also, our matching at low-energy ---to $f_{\pi}$
and $M_{\rho}$--- corresponds to  $N_c \simeq 4.3$ (identifying the factor $l_0 /
g_5^2$ in  (\ref{HEPVA})). To correct these flaws, one can modify the
metric {\footnote{This is different from the approach used in
{\cite{Erlich:2005qh,daRold:2005zs}}.}}. A metric that behaves near the UV
brane as $\frac{z}{l_0} w \left( z \right) - 1 \underset{z \rightarrow
l_0}{\sim} (z-l_0)^{2 d}$ will generate a $1 / Q^{2 d}$ correction to (\ref{HEPVA}).
This introduces new parameters, allowing to impose $N_c = 3$ while preserving
numerical agreements for low-energy quantities. The case $d = 2$ will
represent a gluon condensate. Modifications with $d = 3$, if felt differently
by the vector and axial fields, will describe the $\left\langle \overline{q} q
\right\rangle^2 / Q^6$ term.

\section{Conclusions}

An $\tmop{SU} ( N_f ) \times \tmop{SU} ( N_f )$ Yang-Mills model with a
compact extra-dimension implements some essential properties of the pions and
vector and axial resonances of large-$N_c$ QCD. The model in its simplest form
contains two parameters: the 5D gauge coupling and another parameter related
to the geometry.

The simplicity of our model comes from the way the spontaneous breaking of the
4D global $\tmop{SU} \left( N_f \right) \times \tmop{SU} \left( N_f \right)$
symmetry is implemented: by BCs. As a consequence, the 5D Yang-Mills fields
yield a multiplet of pions descending from the fifth component of the axial
vector field $A_5$, without the need for any other degree of freedom.

Our results can be summarized in four points:
\begin{enumeratenumeric}
  \item We have derived, for a generic metric, the low-energy limit of such a
  model, while at the same time preserving chiral symmetry. This ensures that
  the lagrangian obtained is that of Chiral Perturbation Theory.
  
  \item We have extracted resonance interactions and demonstrated the ensuing soft
  HE behavior. This result is {\tmem{independent of the 5D
  metric}}. It allows to match the behavior of QCD for some
  amplitudes.

  \item $\rho$-meson dominance holds due to the 5D structure. It ensures that low-energy quantities
  give good estimates of the QCD ones, and are {\tmem{not very sensitive to
  the geometry}}. However, it should be noted that the way
  $\rho$-meson dominance  is
  realized depends on the precise form of the metric.

  \item We have derived an analytic expression for the vector and axial two-point
  functions at all momenta. This is done for  AdS$_5$ space, which enables a further
  matching with the partonic logarithms of QCD {\cite{Son:2003et}}.
We also propose modifications of the metric as a means of  improving the matching with
the QCD OPE.
\end{enumeratenumeric}

\section*{Acknowledgments}
We thank Misha Shifman for stimulating exchanges.
JH is supported by
the European network EURIDICE (HPRN-CT-2002-00311) and by the
Generalitat Valenciana (GV05/015).

\end{document}